\documentclass[
	aps,
	prl,
	superscriptaddress,
	onecolumn,
	preprint,
 ]{revtex4-1}
\usepackage{dcolumn,bm,textcomp}
\usepackage{mathtools}
\usepackage{amsfonts}
\usepackage{textcomp}
\usepackage{floatrow} 
\usepackage{graphicx}
\usepackage{mathptmx}
\usepackage{xspace}
\usepackage{type1cm}
\usepackage{multirow}
\usepackage{textgreek}
\usepackage{natbib}
\usepackage{url}
\makeatletter
\renewcommand{\@biblabel}[1]{[#1]\hfill}
\makeatother

\begin{document}
\title{Gate-controlled reversible rectifying behaviour in tunnel contacted atomically-thin MoS$_{2}$ transistor}


\author{Xiao-Xi Li}
\affiliation{Shenyang National Laboratory for Materials Science, Institute of Metal Research, Chinese Academy of Sciences, Shenyang 110016, China, $\&$ School of Material Science and Engineering, University of Science and Technology of China, Anhui 230026, China}

\author{Zhi-Qiang Fan}
\affiliation{State Key Laboratory of Superlattices and Microstructures, Institute of Semiconductors, Chinese Academy of Sciences, Beijing 100083, PR China}

\author{Pei-Zhi Liu}
\affiliation{Key Laboratory of Interface Science and Engineering in Advanced Materials of Ministry of Education, Taiyuan University of Technology, Taiyuan 030024, PR China}

\author{Mao-Lin Chen}
\affiliation{Shenyang National Laboratory for Materials Science, Institute of Metal Research, Chinese Academy of Sciences, Shenyang 110016, China, $\&$ School of Material Science and Engineering, University of Science and Technology of China, Anhui 230026, China}

\author{Xin Liu}
\affiliation{International Center for Quantum Materials, School of Physics, Peking University, Beijing 100871, PR China}
\affiliation{Collaborative Innovation Center of Quantum Matter, Beijing 100871, PR China}

\author{Chuan-Kun Jia}
\affiliation{College of Materials Science and Engineering, Changsha University of Science $\&$ Technology, Changsha, 410114, China}

\author{Dong-Ming Sun}\email{dmsun@imr.ac.cn}
\affiliation{Shenyang National Laboratory for Materials Science, Institute of Metal Research, Chinese Academy of Sciences, Shenyang 110016, China, $\&$ School of Material Science and Engineering, University of Science and Technology of China, Anhui 230026, China}

\author{Xiang-Wei Jiang}\email{xwjiang@semi.ac.cn}
\affiliation{State Key Laboratory of Superlattices and Microstructures, Institute of Semiconductors, Chinese Academy of Sciences, Beijing 100083, PR China}

\author{Zheng Han}\email{vitto.han@gmail.com}
\affiliation{Shenyang National Laboratory for Materials Science, Institute of Metal Research, Chinese Academy of Sciences, Shenyang 110016, China, $\&$ School of Material Science and Engineering, University of Science and Technology of China, Anhui 230026, China}

\author{Vincent Bouchiat}
\affiliation{University of Grenoble Alpes, CNRS, Institut N$\acute{e}$el, F-38000 Grenoble, France}

\author{Jun-Jie Guo}
\affiliation{Key Laboratory of Interface Science and Engineering in Advanced Materials of Ministry of Education, Taiyuan University of Technology, Taiyuan 030024, PR China}

\author{Jian-Hao Chen}
\affiliation{International Center for Quantum Materials, School of Physics, Peking University, Beijing 100871, PR China}
\affiliation{Collaborative Innovation Center of Quantum Matter, Beijing 100871, PR China}

\author{Zhi-Dong Zhang}
\affiliation{Shenyang National Laboratory for Materials Science, Institute of Metal Research, Chinese Academy of Sciences, Shenyang 110016, China, $\&$ School of Material Science and Engineering, University of Science and Technology of China, Anhui 230026, China}


\maketitle


\textbf{Atomically-thin two dimensional semiconducting materials integrated into van der Waals heterostructures have enabled architectures that hold great promise for next generation nanoelectronics. However, challenges still remain to enable their applications as compliant materials for integration in logic devices. Two key-components to master are the barriers at metal/semiconductor interfaces and the mobility of the semiconducting channel, which endow the building-blocks of \textit{pn} diode and field effect transistor. Here, we have devised a reverted stacking technique to intercalate a wrinkle-free boron nitride tunnel layer between MoS$_{2}$ channel and source drain electrodes. Vertical tunnelling of electrons therefore makes it possible to suppress the Schottky barriers and Fermi level pinning, leading to homogeneous gate-control of the channel chemical potential across the bandgap edges. The observed unprecedented features of ambipolar \textit{pn} to \textit{np} diode, which can be reversibly gate tuned, paves the way for future logic applications and high performance switches based on atomically thin semiconducting channel.}

A decade after the first isolation and study of two-dimensional (2D) materials, their atomically-precise integration into van der Waals (vdW) planar heterostructures \cite{Cory_NatNano, Cory_Science} is now forming an outstanding platform for developing novel nanoelectronic devices \cite{Geim_review, Neto_review, Duan_review}. Such platform has been the source of many recent advances in electrical engineering that takes the advantages of the coupling of mono- or few-layered 2D materials such as graphene, hexagonal boron nitride (h-BN) and Transition Metal Dichalcogenides (TMDCs). It has thus far thrived a rich variety of physical phenomena, including Metal Oxide Semiconductor Field Effect Transistors (MOSFETs) \cite{Cory_NatNano}, spintronics memory devices \cite{SpinValve_PRL}, photovoltaics \cite{Duan_review}, and atomically thin superconductors \cite{Iwasa_Science}. Although doping control by an electrostatic gate in those devices has enabled tremendous opportunities, the lack of gapped 2D channel with complementary ($p$ and $n$) polarities has hampered its application in logic units based on the co-manipulation of diodes and field effect transistors, each has been the core of modern electronics. MoS$_{2}$ is among the most studied TMDC compounds for both its outstanding electronics and optoelectronics properties as it combines well-defined bandgap, stability in ambient conditions and relatively high charge carrier mobility. Indeed 2H-type molybdenum disulfide (2H-MoS$_{2}$) has  a thickness-dependent bandgap of 1.3 eV indirect gap $\sim$ 1.9 eV direct gap from bulk down to single layer, respectively \cite{Heinz_PRL}. It therefore holds great promise not only for fundamental studies \cite{Iwasa_Science, Xu_Cui_NatNano, Valley_Hall}, but also for future applications such as high performance FETs and opto-electronics \cite{Kis_NatNano, Nature_BTBT_MoS2, MoS2_1nmGate, Piezo_ZLWang, NC_MoS2, Duan_NN, WeidaHu_AM}. Field effect transistors involving atomically-thin MoS$_{2}$ \cite{Kis_NatNano} as the active channel have enabled original architectures which unlock new features such as sub-thermionic inter-band tunnelling exhibiting unprecedented high sub-threshold swing \cite{Nature_BTBT_MoS2}, or ultra-short gate-length FETs \cite{MoS2_1nmGate}, opening promising pathways for further enhanced integration.

To fulfill the desired performances of CMOS-type logics using MoS$_{2}$ FETs, one of the key (yet evasive) goals has been achieving programmable ambipolar operation (i.e. obtaining easily reconfigurable same-chip $n$- and $p$-doping in MoS$_{2}$ FETs). However, to date, only few experiments \cite{NanoLett_MoO3, Bao_APL, Iwasa_NanoLett} were reported hole transport in MoS$_{2}$, which was achieved through gate dielectric engineering with high gate voltage operation \cite{Bao_APL}, or in an ionic liquid gating environment \cite{Iwasa_NanoLett}. Great efforts have been conducted to pursue ambipolar field effect and further gate tunable rectifying characteristics in MoS$_{2}$ based heterostructures, including MoS$_{2}$ coupled with other materials such as carbon nanotube films \cite{PNAS_Hersam}. Similar effects can also be found in $n$-type TMDCs vdW interfaced with $p$-type TMDCs \cite{NanoScale, ACSnano} or with organic crystal thin films \cite{PCCP}.

  \begin{figure*}[ht]
  \includegraphics[width=0.9\linewidth]{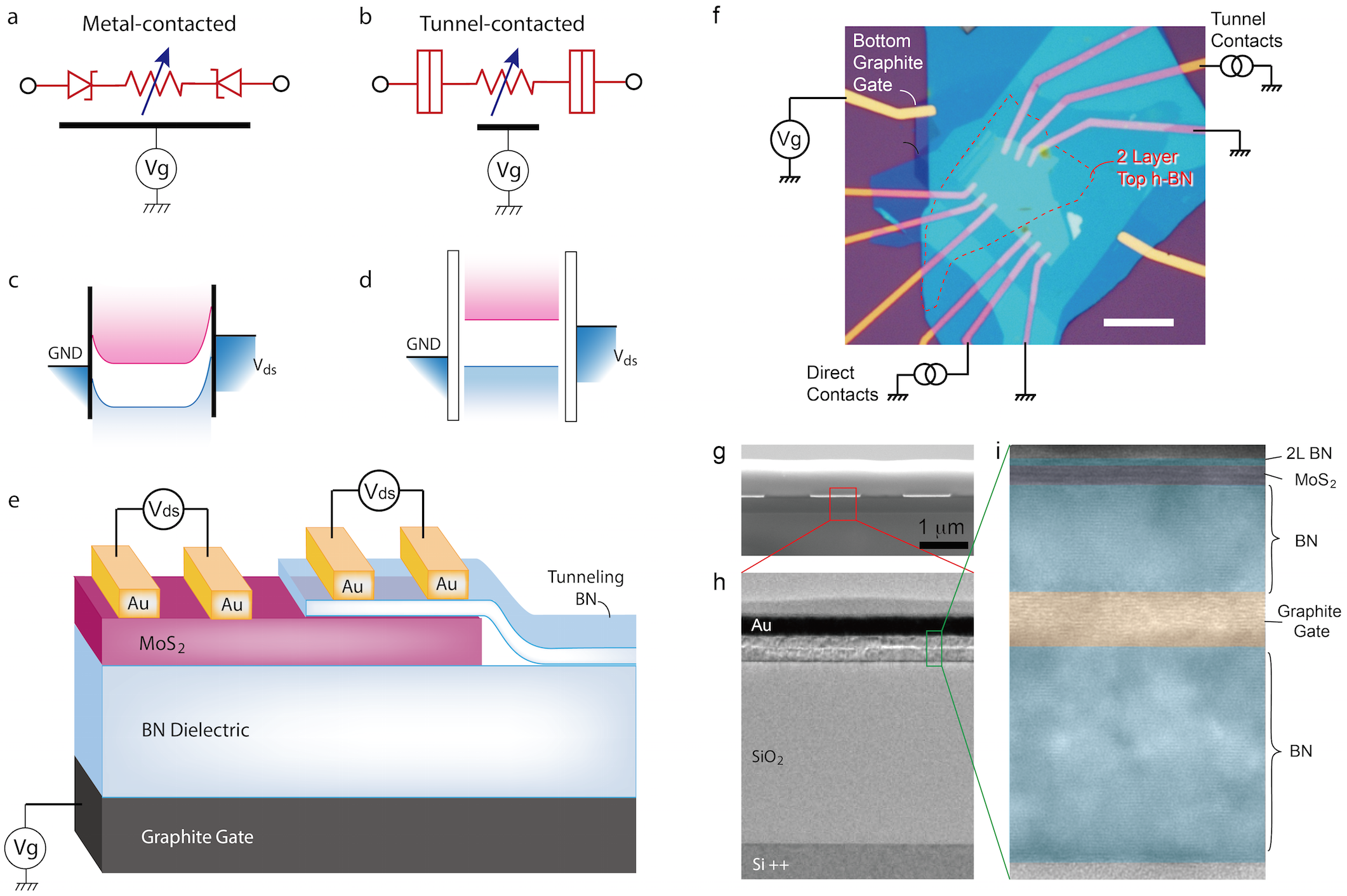} 
  \caption{\textbf{Comparison between metal-contacted and metal/insulator tunnel-contacted MoS$_{2}$ FETs } (a) Schematics of a metal-contacted MoS$_{2}$ film leading to a Schottky barrier field effect transistor (SB-FET). (b) Schematics of a tunnel-contacted MoS$_{2}$ field effect transistor (TC-FET). (c-d) Semiconductor representation of the energy levels respectively for SB-FET and TC-FET showing the absence of band bending in TC-FETs (e) Schematics cross section of the device showing SB-FET and TC-FET side by side on the same MoS$_{2}$ flake. (f) Optical micrograph of a typical TC-FET device. Red dashed line highlights the 2-layered top h-BN, which covers half the MoS$_{2}$. Scale bar is 10 $\mu$m. (g) Scanning electron microscopy (SEM) image of the cross-section of the graphite-gated MoS$_{2}$ vertical tunnel device, with its boxed area zoomed in transmission electron microscopy (TEM) images in (h) and (i).
  }
  \end{figure*}

 \section{h-BN as a ultrathin dielectrics for carrier injection via tunelling}
 
In this work, we demonstrate the design and room temperature operation of FETs based on a tunnel-contacted (TC) MoS$_{2}$ channel. The tunnel barrier insulating layer is implemented  by an ultra-thin capping layer that enables the vertical tunnelling of electrons from the top deposited electrodes. Ultra-thin (one or few monolayer) BN has been identified in the past as an efficient dielectric essential to a number of vertical transport devices, including graphene tunnel transistors \cite{Britnell_NanoLett, Britnell_Science, Mishchenko_NatNano, Eisenstein_NanoLett, Hone_APL_2011, DGG_PRB_2012}, and excitonic super-fluid double layer systems \cite{Leo, Xiaomeng}.

As the few-layers h-BN is used as the top most layer, it assumes the role of an atomically uniform potential barrier, across which electrons are coupled through the tunneling process. For that purpose it is required to be contaminant- and wrinkle-free. Recent results on shot noise measurements in Metal-hBN-metal tunnel junctions confirm that h-BN behaves as an ideal tunnel barrier \cite{shotnoise_tunnelhBN}. 

With the conventional scheme of metal/MoS$_{2}$ contact, Fermi level pinning at the contact interface usually leads to a gate-dependent Schottky barrier (SB) \cite{Kim_pinning}, which results in extra contact resistance that interferes with device performance (Fig. 1a and 1c). Here, by using the reverted vdW stacking method (Suppl. Info.), large-area and wrinkle-free 2-layered h-BN can be inserted between metal contacts and 2D semiconductor channel. We found that the presence of tunnel barriers in the form of 2-layered h-BN perfectly suppresses the SB, and chemical potential of the MoS$_{2}$ layer can be adjusted in a uniform manner across the entire channel, achieving precise electrostatic control of the Fermi level of the 2D layer (Fig. 1b and 1d). Ambipolar field effect at finite source-drain bias, and consequently fully reversible $pn$ to $np$ diodes by gating was obtained.

\section{Results} 
Schematic together with an optical image of a typical TC-device is shown in Fig. 1e-f. Same flake of few-layered MoS$_{2}$ is contacted by normal metal-contacts, and tunnel-contacted electrodes, respectively. Atomic Force Microscopy (AFM) image confirms that devices made by our reverted vdW stacking method exhibit atomically flat top tunnel layer, which is free of wrinkles nor ruptures over 10 $\times$ 10 $\mu$m$^{2}$ area (Suppl. Info.). The cross-sectional transmission electron microscopy (TEM) specimen prepared by focus ion beam (FIB) of the sample in a local area under metal electrodes is shown in Fig. 1g-i. The multi-layered vdW heterostructure can be clearly seen with a 2-layered h-BN on top of few-layered MoS$_{2}$. To improve the gate efficiency and uniformity \cite{Yuanbo_BP}, graphite flakes with thickness of about 4$\sim$6 nm are used as electrostatic gate spaced by a sub-10 nm h-BN under the MoS$_{2}$ channel (Fig. 1i).

  \begin{figure}[ht!]
  \includegraphics[width=0.5\linewidth]{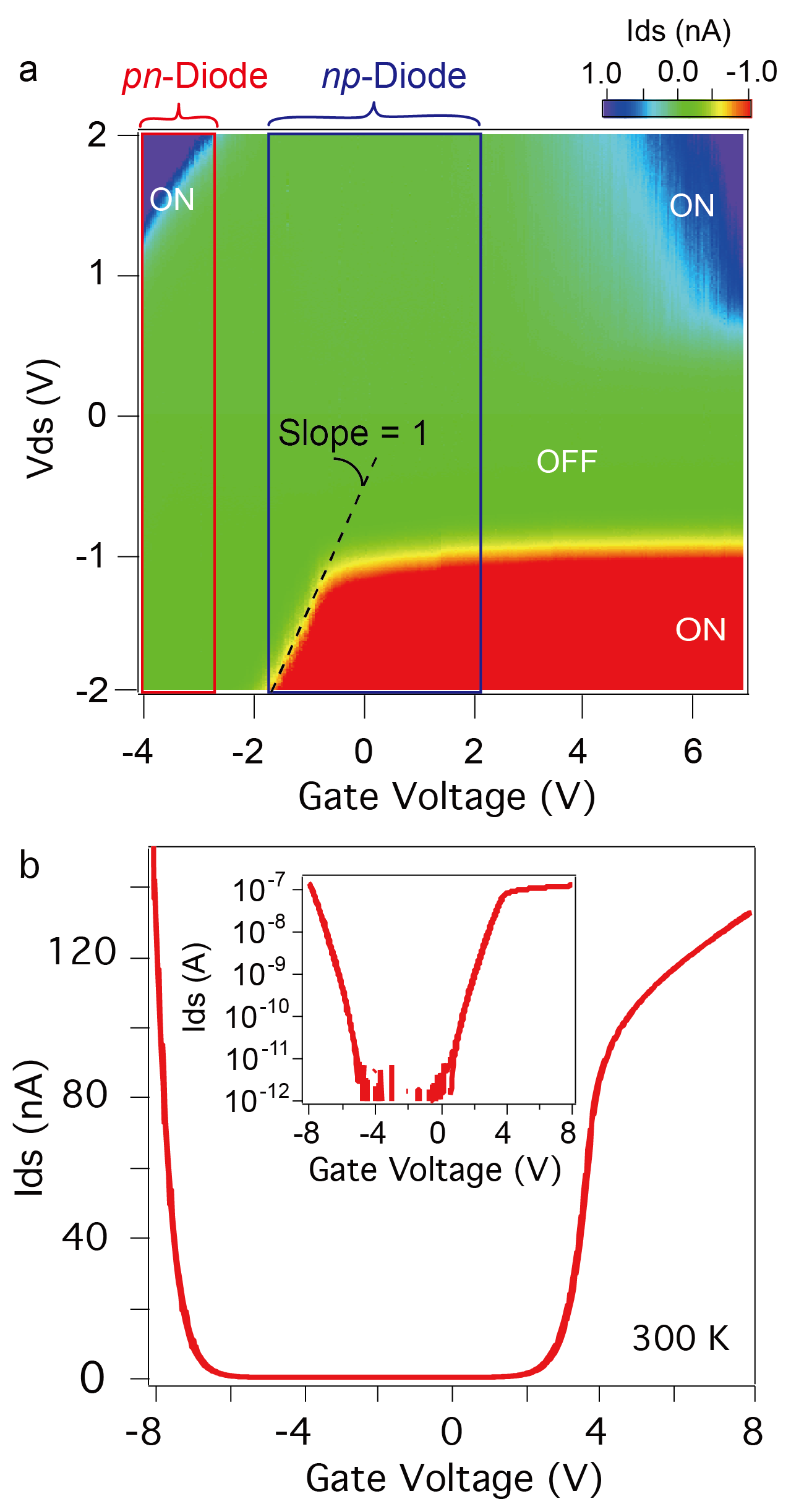}
  \caption{\textbf{Transport characteristics of MoS$_{2}$ TC-FETs.} (a) Color map of output curves ($I_{ds}$ vs $V_{ds}$) at different gate voltages for a typical tunnel-contacted device (room temperature operation). Red and blue-boxed areas highlight the operation range in gate voltage for $pn$ and $np$ diodes, respectively. (b) Typical ambipolar field effect curve at  $V_{ds} = +2$ V measured in samples fabricated by the reverted vdW stacking method. Inset: same data in a semilog plot (more data on other samples in Suppl. Info.). }
  \label{fig:fig2}
  \end{figure}

First, we characterize the MoS$_{2}$ FET with conventional Au (50 nm) electrodes. As shown in Fig. S4, transport measurements of them show typical n-type FET behavior. Color map of $IV$ characteristics at fixed gate voltages indicates ``ON" states at positive and negative bias voltages ($V_{ds}$) on the electron side, while the channel turns off on the hole side. $IV$ curves at fixed $V_{g}$ slightly deviate from linear behavior, while the transfer curves at fixed bias voltage $V_{ds}$ show typical $n$-type unipolar field effect (Fig. S4). These behaviours are standard in MoS$_{2}$ FET, agree with previously reported \cite{Kis_NatNano, Xu_Cui_NatNano, NC_MoS2}.

A striking consequence of the insertion of an ultra-thin h-BN below metal contact is the dramatic change in the color map of $IV$ curves at fixed gate voltages, as shown in Fig. 2a. Instead of the rather symmetric $V_{ds}$ polarization with ``ON" state only seen in the electron side for metal contacted MoS$_{2}$ FET, the vertical tunnel-contacted FET (TC-FET) on the same piece of MoS$_{2}$ flake, as well as in the same gate range, features strongly asymmetric $V_{ds}$ polarization in the whole gate range. Surprisingly, when $V_{ds}$ is larger than a threshold value of about 1 V, the device starts to exhibit ambipolar transfer curves, with ``ON" state observed on both electron and hole sides at source-drain bias above +1 V. A typical  such ambipolar field effect curve is shown in Fig. 2b.

Figure S5 plots the transfer curves cut from Fig. 2a along fixed $V_{ds}$. Largely distinctive from the same piece of normal metal contacted MoS$_{2}$ (Fig. S4), transfer curves for vertical tunnel-contacted MoS$_{2}$ FET (TC-FET) exhibits strong bias asymmetry. The hole side conductivity are comparable to the value on the electron side, reaching a maximum source-drain current $I_{ds}$ of 140 nA. Here the tunnelling current scales positively with respect to the area of tunnel contacts (Fig. S6). We recorded the leakage current from the gate concomitant with transport measurement, to rule out any such influence. It is confirmed that leakage current are limited in a sub 100 pA range, compared to the measured bipolar field effect curve gives $I_{ds}$ up to over 100 nA. To verify the reproducibility of the observed ambipolar FET behavior at finite positive $V_{ds}$, we fabricated multiple samples with similar geometry. All of them show in general similar curves at $V_{ds}= 2$ V, as illustrated in Fig. S7. We also inverted the source drain electrodes and obtained similar behavior within the permit of gate leakage. In the following context, we will focus on the sample in Fig. 2a.

When at negative $V_{ds}$, the transfer curves of TC-FETs become unipolar up to the highest tested gate range (Fig. S5), but still varies from conventional FET. After the turn-on point, $I_{ds}$ initially increases exponentially, then enters a broad saturation plateaux at higher electron doping. It is noteworthy that a shoulder toward saturation is often seen also on the electron side of the transfer curve at positive $V_{ds}$, as indicated in Fig. 2b. We note that recent report shows that a monolayer chemical vapor deposited h-BN spacing layer can diminish Schottky barrier at the metal contact, giving rise to a tripled output current in the transistor \cite{LiaoLei_AM}. However, we didn't see such behavior in our vertical tunnel devices, which may be a result of the less-defected h-BN crystals used in this work.

\begin{figure*}[ht!]
\includegraphics[width=0.9\linewidth]{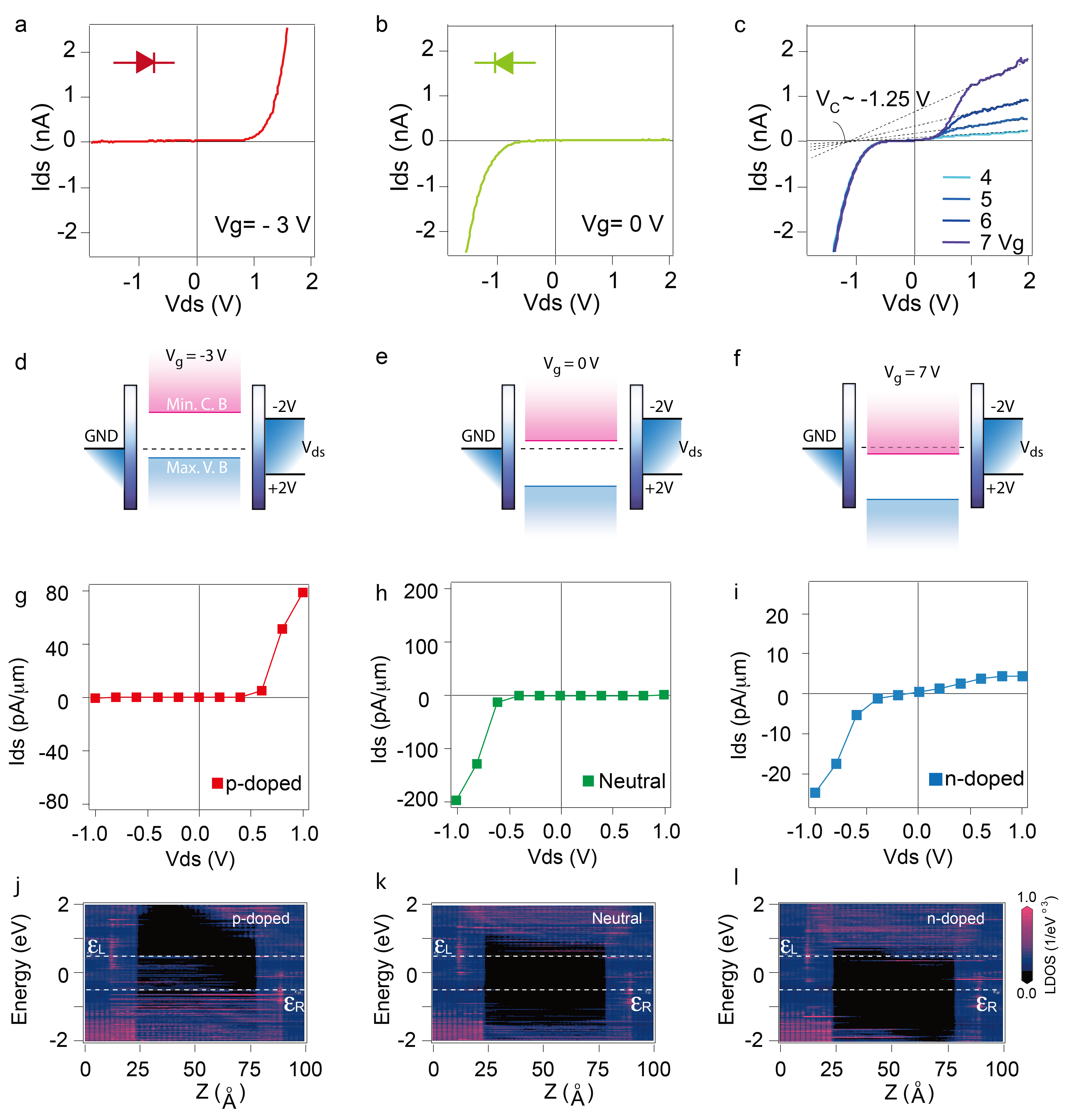}
\caption{\textbf{Room temperature gate-controlled reversible rectifying diode in a TC-FET} (a)-(c) IV curves showing perfect rectifying behavior with reversible polarity characteristics of MoS$_{2}$ TC-FETs. a to c are linecuts of Fig. 2a, with output curves along fixed gate voltages of -3 V, 0 V, and 4-7 V, respectively. While (d)-(f) are the corresponding schematic band alignment pictures. (g-l) Simulations of rectifying characteristics of tunnel contacted MoS$_{2}$ FET. (g)-(i) Simulated $IV$ characteristics of the MoS$_{2}$ vertical tunnel FET at hole-doping, neutral, and electron-doping, respectively. At these corresponding doping level, their simulated PLDOS at $V_{ds}$= +1.0 V are shown in (j)-(l).} 
\label{fig:fig3}
\end{figure*}

To better understand the obtained result in Fig. 2a, we now focus on the line cuts of $IV$ along fixed $V_{g}$. It is found that, at the largest negative gate voltage of about -3 V (all $V_{g}$ in the measurements were pushed to the leakage limit), the output curves behave as typical $pn$ diode with rectification characteristics, and on/off ratio over 10$^{5}$ (Fig. 3a). When gate voltage is brought into the range of -1 V to +2 V, it is seen that the diode behavior is inverted into $np$ type by solely tuning the gate (Fig. 3b). The ``ON" side is now in the negative bias voltage direction, as marked by boxes in Fig. 2a. Upon further doping to the electron side, i.e., at larger positive gate voltages, the output curves gradually shift from the diode behavior into an asymmetric $IV$ with the low bias range following the conventional semiconducting trend, but rather linear at large positive bias. Strikingly, the linear parts can be extrapolated into a single crossing point on the zero-current axis, with a crossing voltage $V_{C}$ of about -1.25 V. This extrapolated crossing point of $IV$ curves is not readily understood and provides food for further experimental and theoretical studies. 

We propose a simple band alignment model to explain the observed behavior of gate-induced switching between $pn$ to $np$ diodes. In conventional metal-contacted MoS$_{2}$ devices, due to the work function mismatch, Schottky barrier forms at the interface of metal and 2D materials, as a result of Fermi level pinning and band-bending near the interface (Fig. 1a and 1c). However, tunnel h-BN in our case overcomes this problem, leading to a relatively free moving conduction and valence bands (Fig. 1b and 1d). At each stage of electrostatic doping in Fig. 3a-c, Fermi level sits at a fixed energy between the minimum of conduction band (CBM) and the maximum of valence band (VBM), respectively. This free band alignment model offers a good description of the $pn$ to $np$ diode inversion in a $V_{ds}$ range of $\pm 2 $V, as illustrated in Fig. 3d-f. Moreover, when Fermi level entering conduction band from the band gap, a slope of unity in $V_{ds}$ vs $V_{g}$ can be extracted in Fig. 2a in the negative $V_{ds}$ regime, indicating a strong energetic coupling of chemical potential from the electrostatic gate to the electronic band in the few-layered MoS$_{2}$ channel. Once the Fermi level enters the valence band, the gate becomes capacitively coupled owing to the large density of states, giving rise to a significantly reduced slope of $V_{ds}$ vs $V_{g}$.

\begin{figure*}[ht!]
\includegraphics[width=0.7\linewidth]{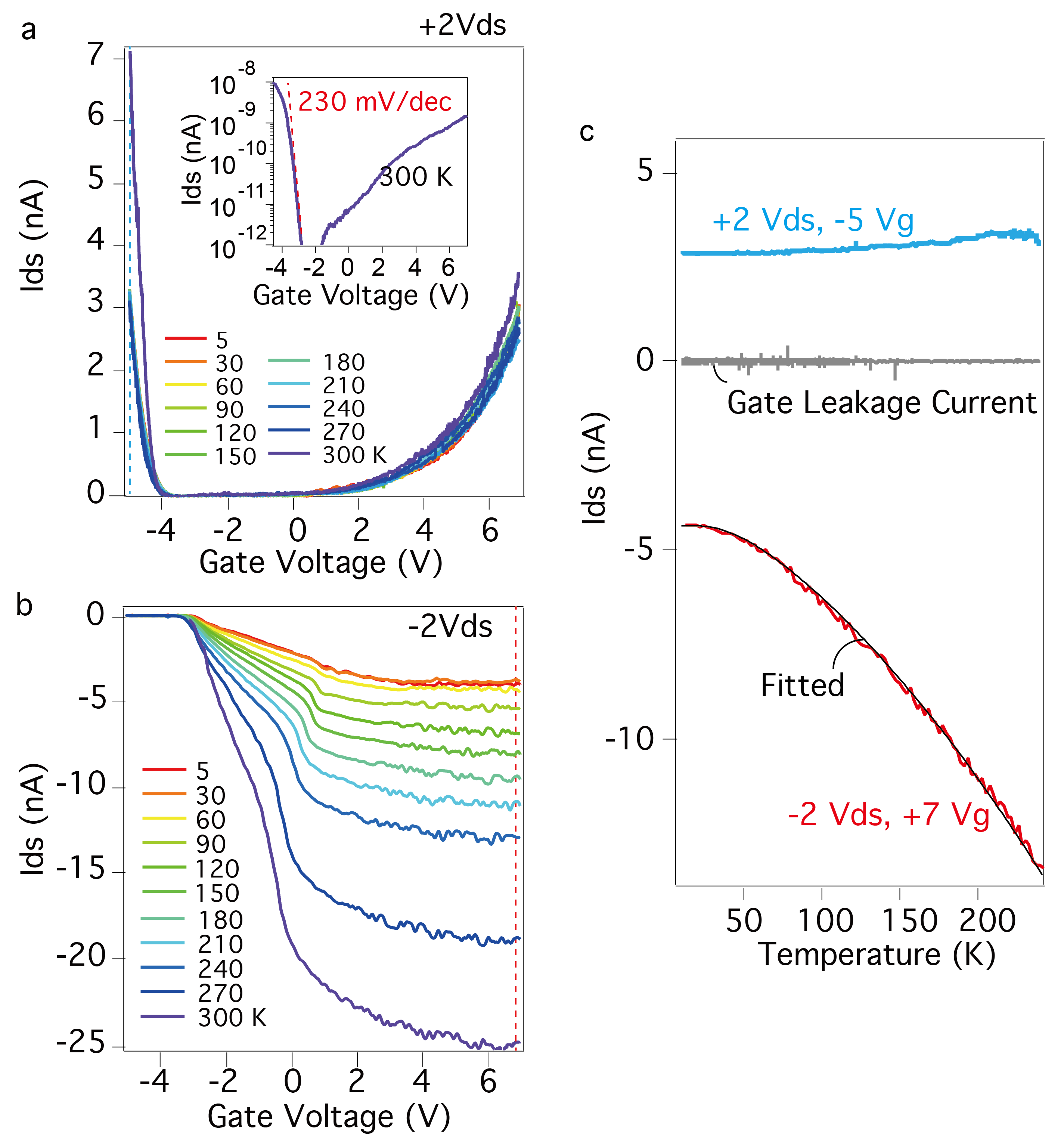}
\caption{\textbf{Temperature dependence of transfer curves in a MoS$_{2}$ TC-FET.} (a)-(b) Transfer curves at different temperatures for the device shown in Fig. 2, at drain source voltages $V_{ds}$=+2 and -2 V, respectively. Inset in (a) is a log scale of the field effect curve. (c) Line traces of temperature dependence of $I_{ds}$ at fixed gate voltage along the blue and red dashed lines in Fig. 5a and 5b, respectively. Gray solid line indicates the gate leakage current during the same measurement. Solid fitting line in Fig. 5c is fitted using eq.(1) in the main text.}

\label{fig:fig4}
\end{figure*}

In the following, we compare the measured data with first principles simulations. For simplicity, we consider the simplest scenario of monolayer MoS$_{2}$ tunnel device with a channel length of about 6 nm (more detail in Suppl. Info.). Compared to Fig. 3a-c, first principles calculations based on the simplified model give qualitative agreement with  experimental observations. As shown in Fig. 3g-i, the 2-layered h-BN tunnel-contacted MoS$_{2}$ FET in our calculated model shows $pn$, $np$, and asymmetrical full pass rectifying characteristics at hole-doping, neutral, and electron-doping, respectively. Their corresponding projected local density states (PLDOS) at $V_{ds}$= +1.0 V are shown in Fig. 3j-l. One can see in the LDOS that the effective transmission forbidden region $\Delta$ in tunnel-contacted device is about 2.5 eV, which is largely enhanced due to the existence of h-BN tunnel barrier ($\Delta \sim$ 1.8 eV in normal contacted device, shown in Fig. S9). The simulated results echo our hypothesis of free band alignment model in Fig. 3d-f. Fermi level pinning in metal contacted devices are suppressed by ultra thin tunnel contact, resulting in the observed finite-bias ambipolar field effect, as well as gate tunable rectifying characteristics with multiple operation states.

\begin{figure*}[ht!]
\includegraphics[width=0.5\linewidth]{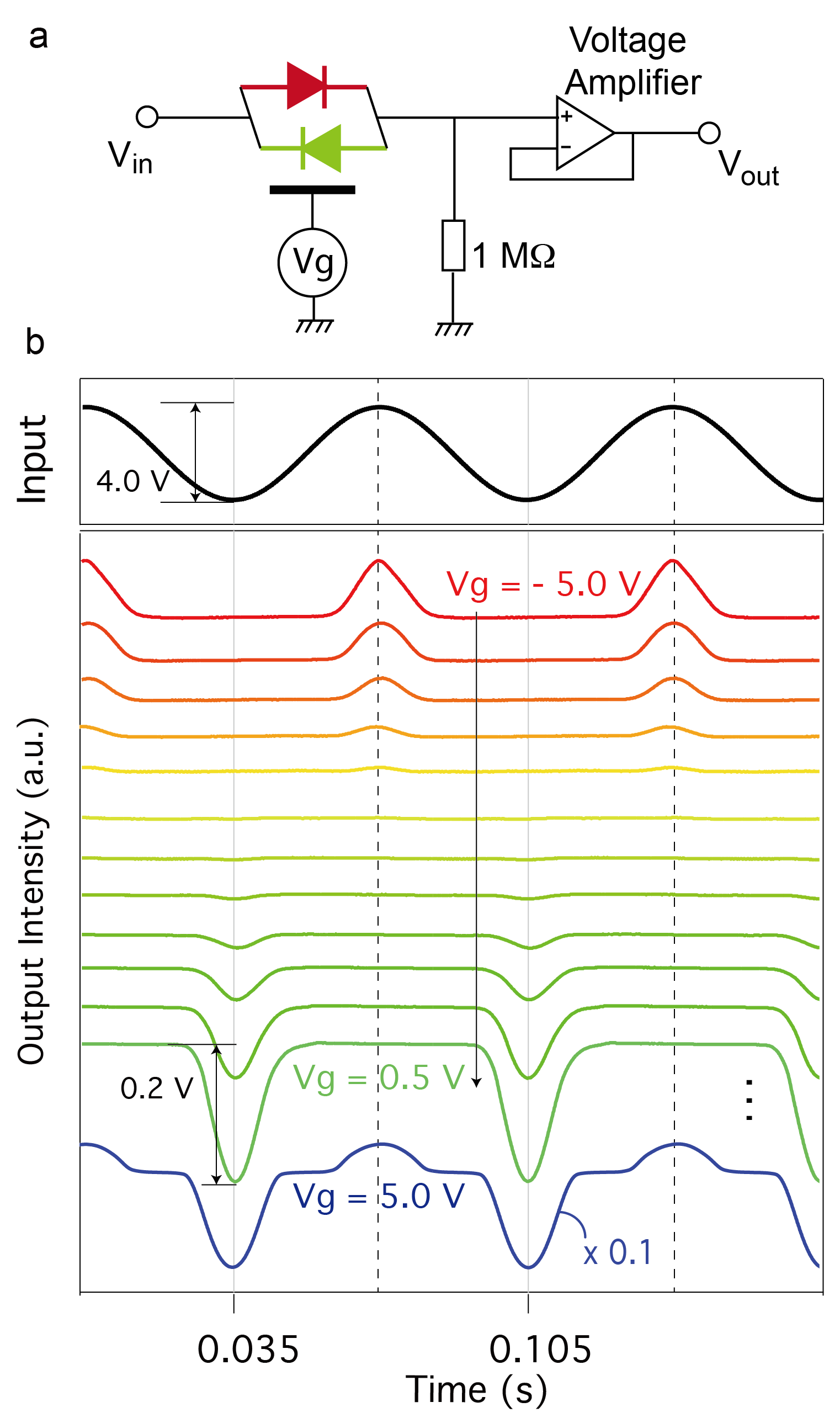}
\caption{ \textbf{Reversal rectification of an analogue harmonic signal in MoS$_{2}$ TC-FET.} (a) Schematics of the gate-control rectifier device placed in a measurement and biasing circuit. The MoS$_{2}$  TC-FET is symbolized as a polarity-switchable  diode. (b) Input (harmonic signal $\sim$ 13 Hz) and output waves of the gate-controlled diode. A $\pi$ phase shift, together with multiple states of output level (e.g. $pn$ diode, ``OFF", $np$ diode, and full pass), in the rectified output wave can be seen via gating. Each measured curve was averaged over 150 recorded traces.}
\label{fig:fig5}
\end{figure*}

It is of fundamental interest to study the temperature dependence of tunnelling current in the MoS$_{2}$ TC-FETs. Figure 4a-b plots the transfer curves of the same device in Fig. 2a, with $V_{ds}=\pm 2$ V at different temperatures from 300 K down to 5 K. It can be seen in Fig. 4a that bipolar transfer curves at $V_{ds}=+2$ V show very weak temperature dependence. A plot of the transfer curve at 300 K is plotted in the inset of Fig. 4a, the sub-threshold swing is extracted on the hole side to be about 230 mV/dec, lower than the 60 mV/decade theoretical limit \cite{Sze}. On the contrary, at $V_{ds}=-2$ V, the transfer curves show rather strong temperature dependence (Fig. 4b), with the $I_{ds}$ decreasing upon lowering the temperature. Single traces of $I_{ds}-T$ monitored at $+2 V_{ds}/-5V_{g}$, and $-2 V_{ds}/+7V_{g}$ are plotted in Fig. 5c, colors are picked according to the dashed lines in Fig. 5a and 5b, respectively. The negatively source-drain biased $I_{ds}-T$ curve at +7V$_{g}$ (red curve) can be fitted by a phonon-assisted tunnelling model \cite{Pho_Ass_Tunnel}:

\begin{equation} 
I \propto \frac{eE}{(8m^{*}\epsilon_{T})^{1/2}}[\Omega-\gamma]^{1/2}[1+\gamma ^{2}]^{1/4}\exp\left\{-\frac{4}{3}\frac{(2m^{*})^{1/2}}{eE\hbar}\epsilon_{T}^{3/2} [\Omega-\gamma]^{2} [\Omega+\frac{1}{2}\gamma]   \right \}
\end{equation}

\noindent where $\gamma=\alpha\sqrt{2m^{*}\epsilon_{T}}\frac{(\hbar \omega)^{2}}{eE}(2[\exp(\hbar \omega/k_{B}T)-1]^{-1}+1)$, and $\Omega=(1+\gamma ^{2})^{1/2}$, with $\alpha$ being a fitting parameter, $E$ the electrical field strength, $\epsilon_{T}$ the tunnel energetic depth, $m^{*}$ the effective electron mass, $\hbar \omega$ the energy of the phonon taking part in the tunneling process, $e$ and $k_{B}$ the element charge and Boltzmann's constant, respectively. Using an effective mass of about 0.018 $m_{e}$ \cite{Taiwan_me}, the best fit in the black solid line Fig. 4c gives $\epsilon_{T} = 0.6$ eV and $\hbar \omega \sim 11 $ meV.

Finally, as a proof of principle for realizing gate-tunable rectifier in the MoS$_{2}$ TC-FET, we used a simple diode circuit with load resistor of 1 M$\Omega$ and output to a 100 M$\Omega$ impedance voltage amplifier (1$\times$ amplification was used in the measurement), as illustrated in the schematics in Fig. 3g. As seen in Fig. 3h, when a sinusoidal wave is input in the MoS$_{2}$ tunnel FET, output wave starts from a positively rectified half wave in the largest hole doping side, and can be first gate tuned into an intermediate ``OFF" state, followed by a negatively rectified half wave in the electron doping side. Further electron doping recovers both positive and negative half output wave, with different amplitude. This gate-tunable rectification inversion with a $\pi$ phase shift phenomenon, together with multiple states of output level (e.g. $pn$ diode, ``OFF", $np$ diode, and full pass), has not been reported before, and can be of great use in future gate-tunable logic circuits with atomically thin conduction channels. It is noteworthy that in a device directly fabricated on SiO$_{2}$, we obtained a cut-off frequency in such MoS$_{2}$ tunnel FET of about 20 kHz when the Si gate is heavily doped (Fig. S8).

In conclusion, we have developed a reverted vdW stacking method for high yield fabrication of resist-free pristine van der Waals heterostructure with ultra-thin top layer. This method itself opens new routes to a number of applications such as scanning tunneling microscope on pristine 2D materials supported by another, as well as the high quality spacing layer for tunneling electrodes. Using this technique, we have demonstrated a vertical tunnel-contacted MoS$_{2}$ transistor, in which suppression of band-bending and Fermi level pinning is realized. The so called Tunnel-Contacted Field Effect Transistor hense gives rise to gate tunable rectification with fully reversible $pn$ to $np$ diode, leading to multiple operation states of output level (e.g. positive-pass, ``OFF", negative-pass, and full-pass). The observed ambipolar field effect at finite positive $V_{ds}$ shows on/off ratio up to 10$^{5}$ in such MoS$_{2}$ FETs, with an output current reaching the order of  100 nA on both electron and hole sides. We proposed a free-moving band alignment model to explain the behavior of the MoS$_{2}$ tunnel FET, which is further qualitatively supported by a simplified first principles simulation model. This work paves the way for future application in gate-tunable logic devices with atomically thin semiconducting channels.

\section{Methods}

In order to have resist-free pristine van der Waals heterostructures, one of the limitations is its stacking sequence: a thick enough h-BN has to be picked up first by polymer (Propylene-Carbonate, PPC, for example) to serve as a top layer. When the top layer is too thin (less than 3 layers), ruptures and wrinkles increase significantly, thus reduce the quality of the final device. We solved this problem by developing a reverted vdW stacking method: Few-layered MoS$_{2}$ is sandwiched by a thick ($\sim$10 nm) BN (crystals from HQ Graphene) and thin (1-3 layer) BN, respectively, with the resulted top later picked up lastly (see SI). When the whole stack is collected, the PPC stamp will be flipped upside down, peeled off with care from the PDMS substrate, and slowly landed onto a hot plate of about 100 $^{o}$C (see SI). At this stage, the stack will be ``floating" on the PPC film, which can be completely evaporated in a vacuum annealer at 350 $^{o}$C for around 20 min. Followed by standard lithography and metallization. MoS$_{2}$ flake is half covered by 1$\sim$3 layer h-BN, and Au electrode with thickness of 20 nm is deposited onto the stack, forming conventional direct contacts and tunnel contacts, respectively. Electronic transport was measured on a Cascade probe station at room temperature, and in a Quantum Design PPMS system with a home-made sample probe interfaced with external measurement setup at low temperatures, respectively. 

The device simulations in this work are carried out by using the first-principles software package Atomistix ToolKit, which is based on density-functional theory in combination with the non-equilibrium Green’s function \cite{GreenFunction}. The exchange-correlation potential is described by the local density approximation (LDA) and the wave function is expanded by the Hartwigsen-Goedecker-Hutter (HGH) basis for all atoms. The real space grid techniques are used with the energy cutoff of 150 Ry in numerical integrations. The geometries are optimized until all residual force on each atom is smaller than 0.05 eV$\AA^{-1}$. The current can be calculated by the Landauer formula\cite{Landauer}:

\begin{equation}
I(V_{ds})=\frac{2e}{h}\int T(E,V_{ds})[f_{S}(E,V_{ds})-f_{D}(E,V_{ds})]dE 
 \end{equation}
 
 Here, $V_{ds}$ is the bias voltage between the drain and the source, $T(E,V_{ds})$ is the transmission coefficient, $f_{S}(E,V_{ds})$ and $f_{D}(E,V_{ds})$ are the Fermi-Dirac distribution functions of the source and drain. The transmission coefficient $T(E,V_{ds})$ as a function of the energy level $E$ at a certain $V_{ds}$ can be calculated by the formula: 
 
 \begin{equation}
T(E,V_{ds})= Tr[\Gamma _{S}(E)G^{R}(E)\Gamma _{D}(E)G^{A}(E)]
  \end{equation}
  
where $G^{R}(E)$ and $G^{A}(E)$ are the advanced and retarded Green’s functions of the scattering region, respectively.

\section{\label{sec:level1}ACKNOWLEDGEMENT}
This work is supported by the National Natural Science Foundation of China (NSFC) with Grant 11504385 and 51627801, and is supported by the National Basic Research Program of China (973 Grant Nos. 2013CB921900, 2014CB920900). X. W. Jiang acknowledges supports by the NSFC Grant 11574304, Chinese Academy of Sciences-Peking University Pioneer Cooperation Team (CAS-PKU Pioneer Cooperation Team), and the Youth Innovation Promotion Association CAS (grand 2016109). D.M. Sun thanks the NSFC grant 51272256, 61422406, 61574143, and MSTC grant 2016YFB04001100. J.H. Chen thanks the NSFC Grant 11374021. Z.D. Zhang acknowledges supports from the NSFC with grant 51331006 and the CAS under the project KJZD-EW-M05-3. V. Bouchiat acknowledges support from the EU FP7 Graphene Flagship (project no. 604391), J2D project grant (ANR-15-CE24-0017) from Agence Nationale de la Recherche (ANR), and the Hsun Lee Award program of the Institute of Metal Research, CAS. The authors are grateful for helpful discussions with Prof. Benjamin Sac$\acute{e}$p$\acute{e}$, Prof. Vasili Perebeinos, and Prof. Ji Feng.


\bibliographystyle{naturemag}

\clearpage
\renewcommand{\figurename}{Fig.S.}
\setcounter{figure}{0}

{\centering\subsection{\large Supplementary Information for\\
``Gate-controlled reversible rectifying behaviour in tunnel contacted atomically-thin MoS$_{2}$ transistor''}

By  Xiaoxi Li, \textit{et al.}

\par}

\vspace{10 mm}

\subsection{\normalsize 1. Material and Methods}

It is known that in order to have resist-free pristine van der Waals heterostructures [Nature \textbf{499}, 419 (2013); Science \textbf{353}, 461 (2016); Nat. Nanotech. \textbf{5}, 722 (2010).], one of the limitations is its stacking sequence: a thick enough h-BN has to be picked up first by polymer (Propylene-Carbonate, PPC, for example) to serve as a top layer. Further layers, including various layered 2D materials, will be dry-picked one after another by van der Waals force with usually all smaller in size compared to the top one [Science \textbf{342}, 614 (2013)]. Modified recipe (with Polycarbonate, PC) is then reported to be more versatile in terms of arranging the stacking sequence, disregarding the size of the to-be-stacked flakes [Appl. Phys. Lett. \textbf{105}, 013101(2014)]. However, when the top layer is too thin (less than 3 layers), ruptures and wrinkles increase significantly, thus reduce the quality of the final device. In particular, PC leaves more contaminations than PPC, when it is in need of flat, ultra-thin, and pristine stack, the conventional stacking technique fails, as indicated in Fig.S1.

\begin{figure*}[h!]
\includegraphics[width=0.75\linewidth]{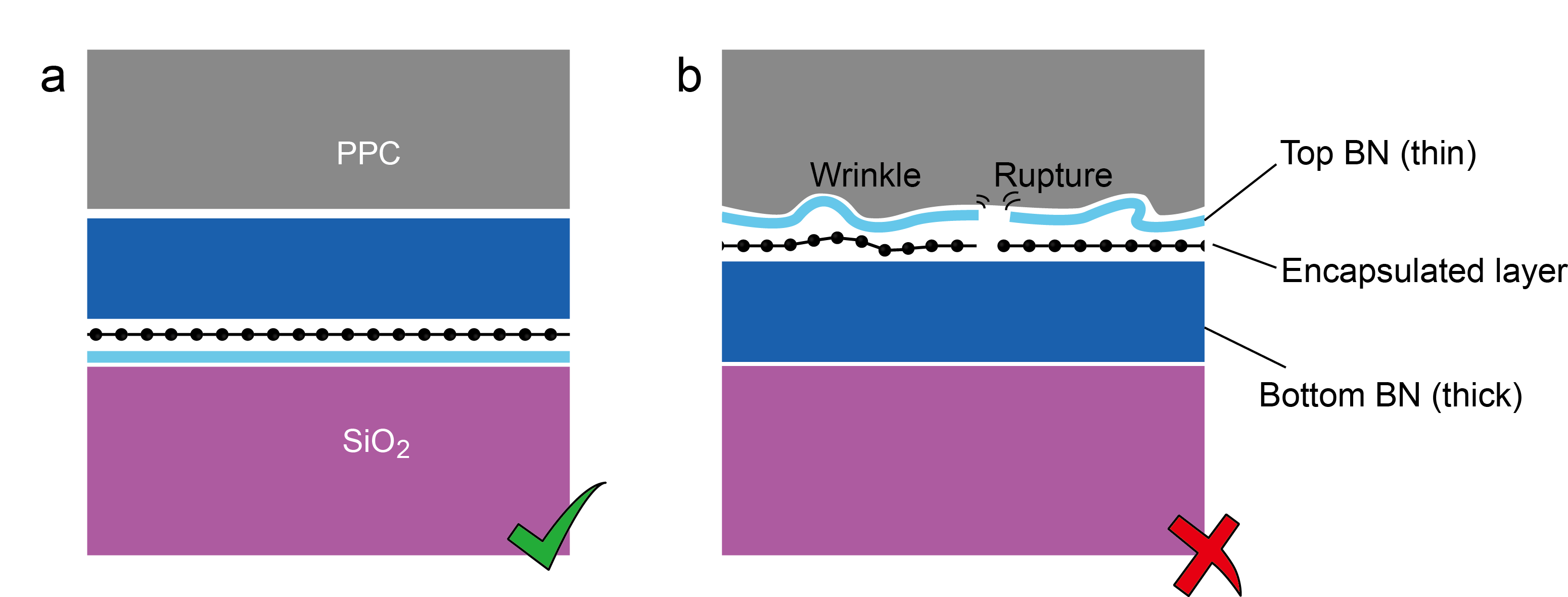}
\caption{Schematics of conventional VdW stacking process with thin layer as bottom (a) and top (b) layer, respectively. When top layer is ultra thin, chances of having wrinkles and ruptures increase significantly (c-d).}
\label{fig:figS1}
\end{figure*}

In this work, we developed a high sample yield reverted van der Waals (vdW) deposition technique, thorough which vdW stacking with ultra-thin pristine top layer can be achieved readily. We first pick up raw layered materials in a reverted order as to the aimed final stack. All layers are collected by the first relatively thick (15-20 nm) h-BN (Fig.S2a) to avoid direct contact with PPC.

\begin{figure*}[h!]
\includegraphics[width=0.8\linewidth]{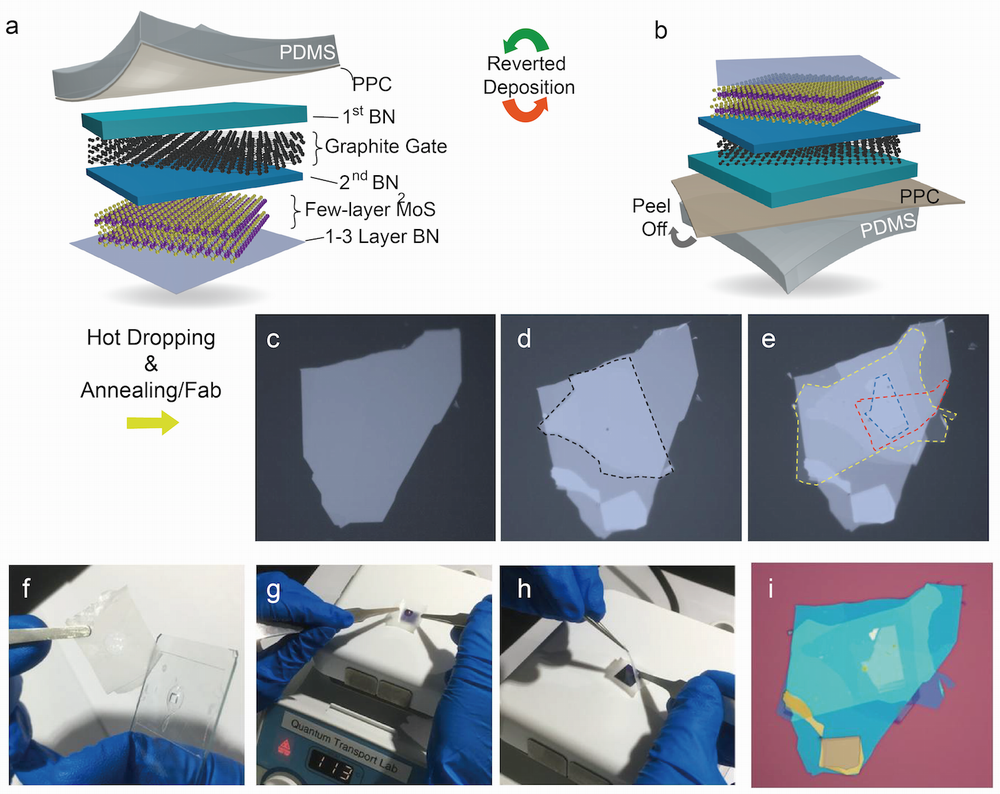}
\caption{(a-b) Cartoon illustrates in an art view of the workflow of reverted vdW staking method for fabricating vertical hetero-structures with high quality wrinkle-free ultra-thin top layer. (c) The top layer of the first thick h-BN picked up by PPC. Subsequent layers will be dry-picked one after another by van der Waals force. (d) A graphite gate with thickness of about 4$\sim$6 nm, to improve the gate efficiency and uniformity. The graphite flake is highlighted with black dashed line. (e) A few-layered MoS$_{2}$ which is sandwiched by a thick ($\sim$10nm) h-BN and thin (1-3 layer) h-BN, respectively. The three flakes (thick h-BN, MoS$_{2}$ and thin h-BN) are highlighted with yellow, blue and red dashed line. (f) PPC with the vdW stack is flipped upside down, and peeled off with care from the PDMS substrate. (g) Landing the peeled PPC (with vdW stack facing up) on a hot SiO$_{2}$/Si substrate, at about 100 $^{o}$C. (h) By removing surrounding Scotch tape, the stack will be left on hot wafer and ``float" on the PPC film. (i) The final stack when underneath PPC was completely removed by vacuum annealing at 350 $^{o}$C for 20 min.}
\label{fig:figS2}
\end{figure*}

Figure S2 illustrates a typical workflow of our process. Opposed to the often-applied stacking method in a top-down sequence [Science \textbf{342}, 614 (2013)], we first pick up raw layered materials in a reverted order (bottom-up sequence) as the aimed final stack. A graphite gate with thickness of about 4$\sim$6 nm is used, to improve the gate efficiency and uniformity, as shown in Fig.S2a-b. Few-layered MoS$_{2}$ is sandwiched by a thick ($\sim$10 nm) BN and thin (1-3 layer) BN, respectively, with the resulted top later picked up lastly (Fig. S2a-b). When the whole stack is collected (Fig. S2c-e), the PPC stamp will be flipped upside down, peeled off with care from the PDMS substrate, and slowly landed onto a hot plate of about 100 $^{o}$C (Fig. S2f-h). At this stage, the stack will be ``floating" on the PPC film, which can be completely evaporated in a vacuum annealer at 350 $^{o}$C for around 20 min. Typical optical image of such final stack with ultra-thin top layer is shown in Fig. S2i.

\begin{figure*}[h!]
\includegraphics[width=0.55\linewidth]{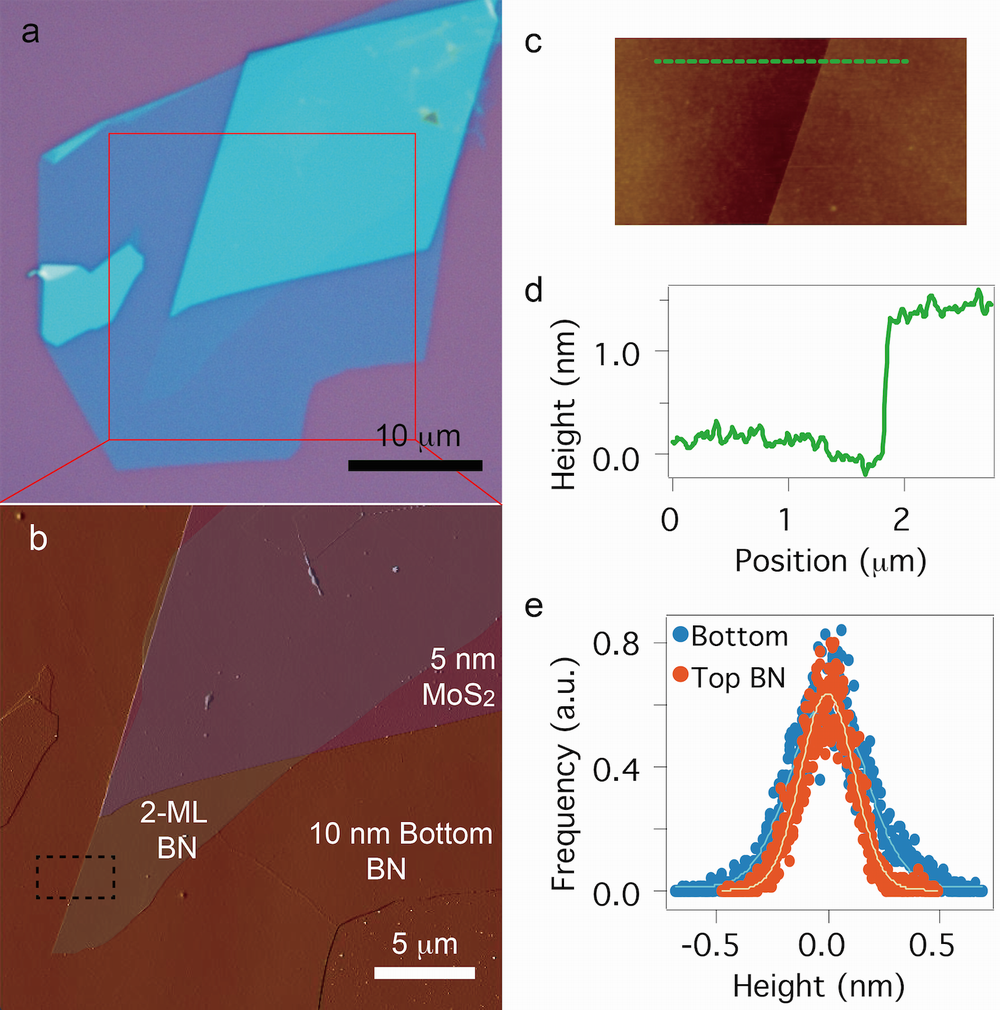}
\caption{(a) Optical micrograph of a typical vertical vdW heterostructure, with 2-layered h-BN on top. Red-boxed area in (a) is scanned in (b) with Atomic force micrograph (AFM) amplitude error image of a typical stack, with a 2-layered h-BN as top layer. (c) Morphology image of black-boxed area in b, with a green dashed line height profile plotted in (d). (e) A statistics of roughness in top and bottom h-BN in the area shown in (c), respectively. Roughness of top-BN is smaller than 0.2 nm.}
\label{fig:figS3}
\end{figure*}

Followed by standard lithography and metallization, the final device is indicated in the schematic in Fig. 1e-f in the main text. MoS$_{2}$ flake is half covered by 2-layered h-BN, and 20 nm thick Au electrodes are deposited onto the stack, forming conventional direct contacts and tunnel contacts, respectively. Atomic force micrograph scan confirms that devices made in this method are of high quality top tunnel layer with roughness smaller than 0.2 nm, and without obvious wrinkles nor ruptures over 10 $\times$ 10 $\mu$m$^{2}$ area (Fig. S3).

\subsection{\normalsize 2. Conventional metal contacted MoS$_{2}$ FETs.}

\begin{figure*}[h!]
\includegraphics[width=0.65\linewidth]{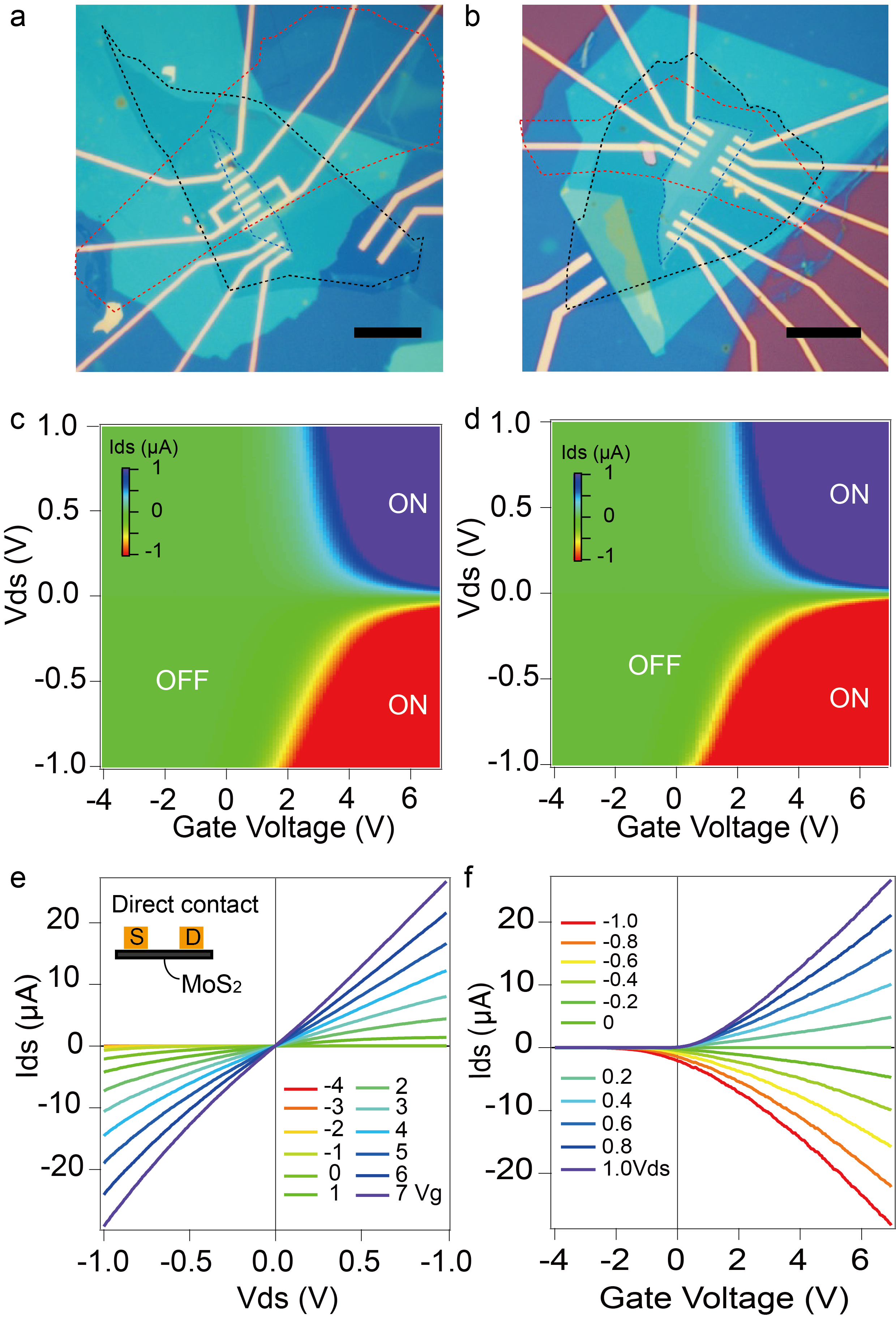}
\caption{(a)-(b) Optical micrographs of two measured devices. Graphite, MoS$_{2}$ and top BN are highlighted with black, blue, and red dashed line, respectively. Scale bars are 10 $\mu$m. (c)-(d) Color map of output curves ($I_{ds}$ vs $V_{ds}$) at different gate voltages for the corresponding direct metal-contacted MoS$_{2}$2 FETs in (a-b), respectively. (e)-(f) Line cuts in (c)-(d), with output curves along fixed gate voltages, and transfer curves ($I_{ds}$ vs $V_{g}$) along fixed bias voltage, respectively.}
\label{fig:figS4}
\end{figure*}

We characterized multiple samples of MoS$_{2}$ FET with conventional Au (50 nm) electrodes. As shown in Fig.S4, the typical transport behavior of n-type MoS$_{2}$ FET can be seen. Color map of $IV$ curves as a function of gate voltages in Fig. S4c-d indicates ``ON" states at positive and negative source-drain bias voltages ($V_{ds}$) at the electron side, with the hole side all turned off. Line cuts of $IV$ along fixed $V_{g}$ are plotted in Fig. S4e, as most of the curves slightly deviated from linear behavior, while conductance increases with increasing gate voltage. Fig. S4f plots the line cuts of transfer curves at fixed bias voltage $V_{ds}$. One can see typical $n$-type unipolar field effect of directly metal-contacted devices. Multi-samples confirm the same behavior, which agrees with the works previously reported [Nat. Nanotech. \textbf{6}, 147 (2011); Nat. Nanotech. \textbf{10}, 534 (2015); Nat. Comm. \textbf{3}, 1011 (2012)].

\subsection{\normalsize 3. Tunnel contacted MoS$_{2}$ FETs.}

\begin{figure*}[h!]
\includegraphics[width=0.7\linewidth]{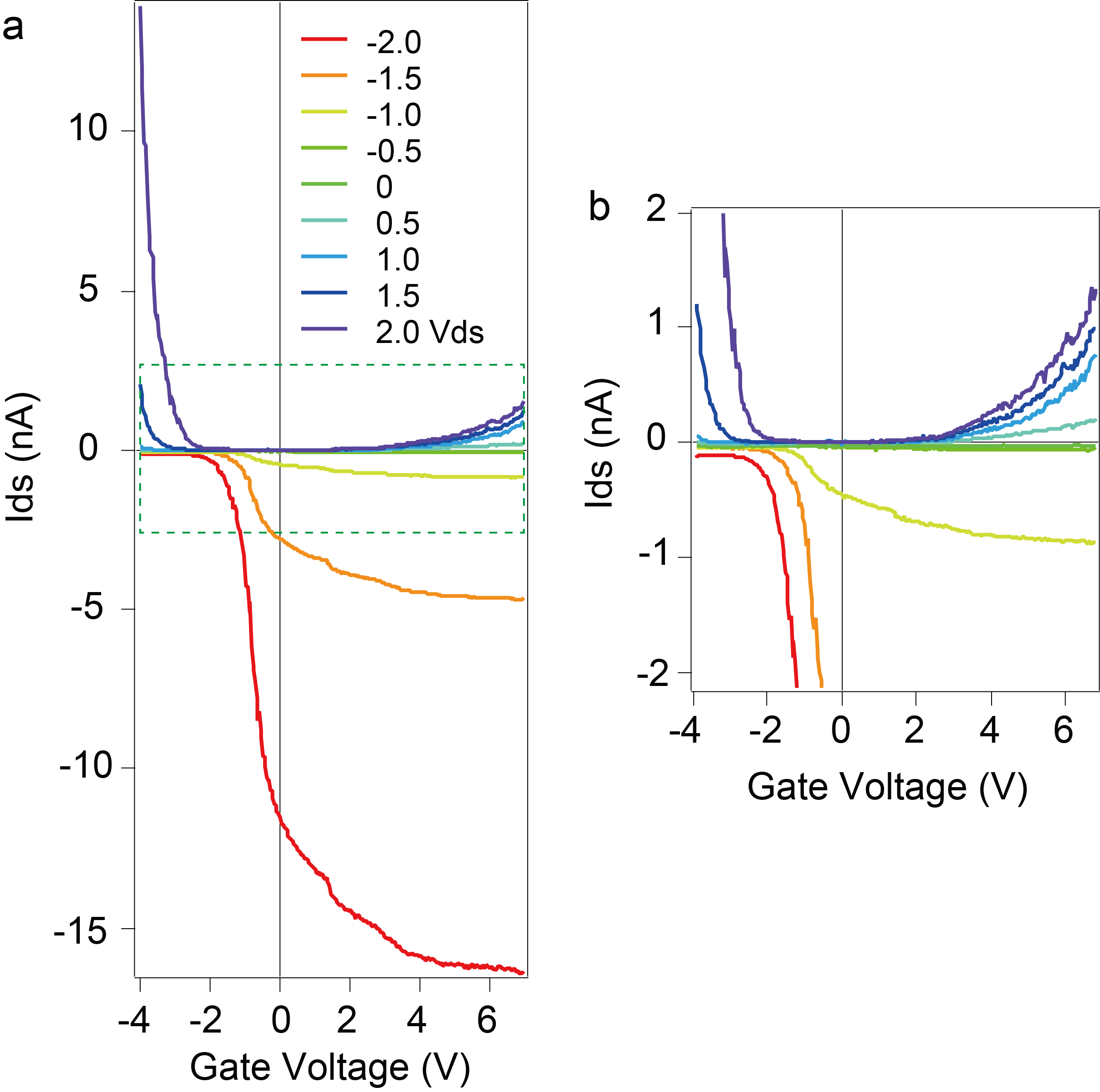}
\caption{Transfer curves cut from Fig. 2a in the main text along fixed bias voltages $V_{ds}$. (b) is a zoomed-in view in the green dashed box in (a).}
\label{fig:figS5}
\end{figure*}

Figure S5 illustrates the line cuts of transfer curves at fixed bias voltage $V_{ds}$. Instead of the rather symmetric $V_{ds}$ polarization with ``ON" state only seen in the electron side for metal contacted MoS$_{2}$ FET (Fig. S4), the MoS$_{2}$ TC-FET on the same piece of MoS$_{2}$ flake, as well as in the same gate range, show strongly asymmetric $V_{ds}$ polarization in the whole gate range. Interestingly, when the tunnel bias voltage is larger than a threshold value of about +1 V, the device start to exhibit ambipolar transfer curves, with the hole side conductivity comparable, sometimes even larger than that of the electron side.

\subsection{\normalsize 4. Contact-area dependence of tunnel current and ambipolar FET.}

\begin{figure*}[h!]
\includegraphics[width=0.75\linewidth]{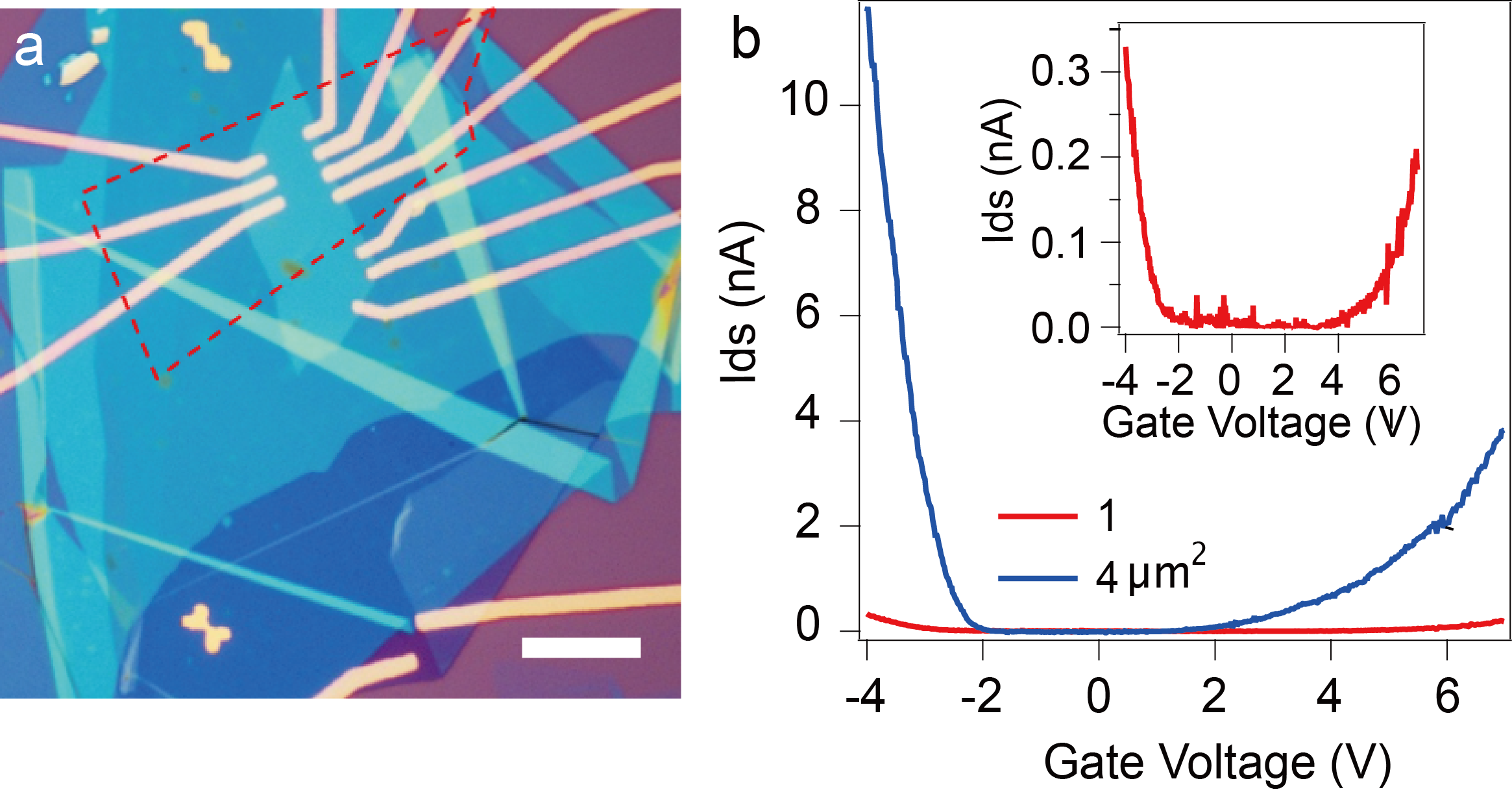}
\caption{(a) Optical micrograph of the device. The top tunnel h-BN is highlighted with red dashed line. Scale bar is 5 $\mu$m. (b) The $I_{ds}-V_{g}$ curves of left tunnel electrodes (blue solid line) and right tunnel electrodes (red solid line) in (a) along +2V bias voltage $V_{ds}$. The inset of (b) is the zoomed-in image of the red solid line in (b).}
\label{fig:figS6}
\end{figure*}

\begin{figure*}[h!]
\includegraphics[width=0.5\linewidth]{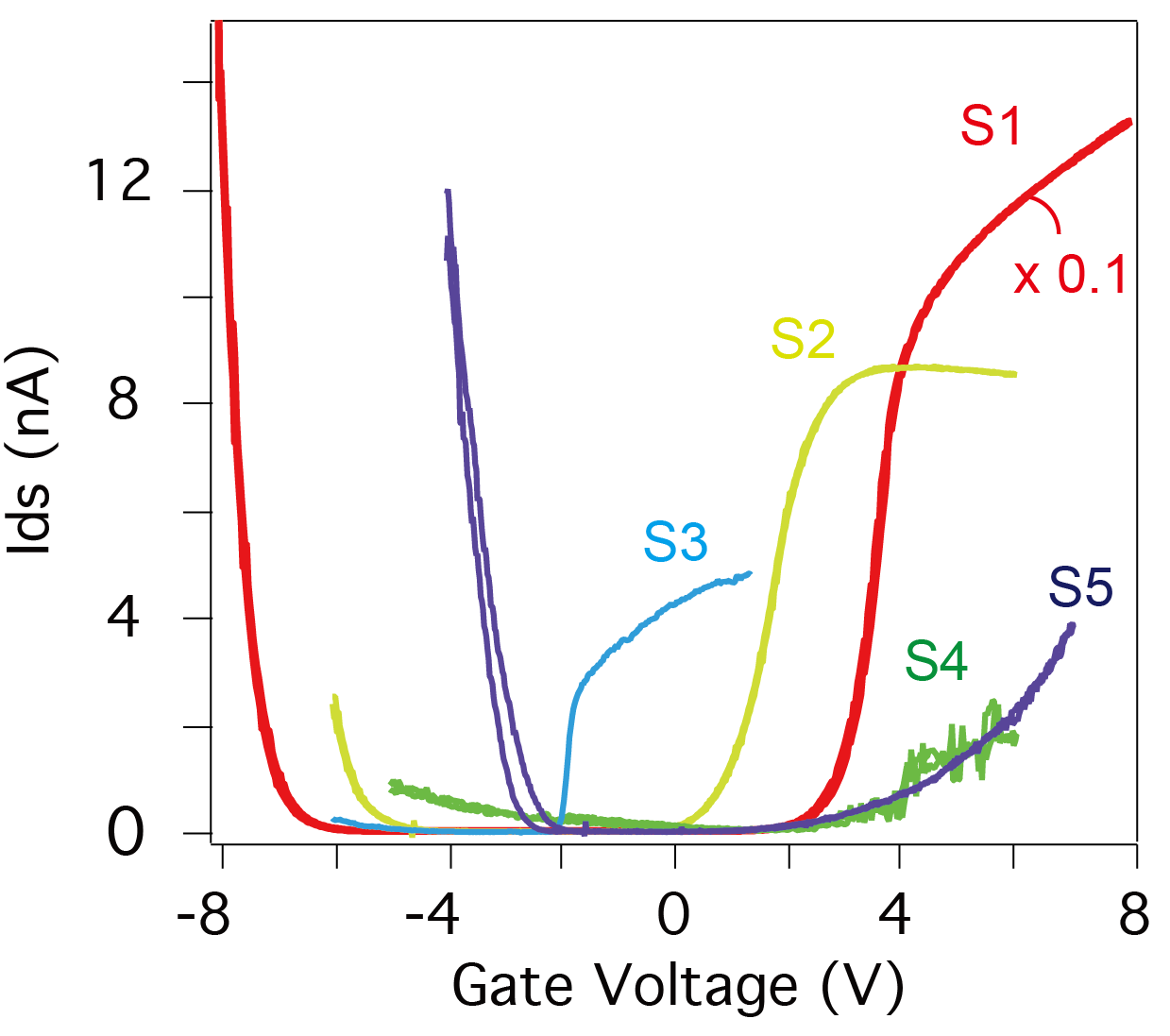}
\caption{Ambipolar field effect curves observed in different MoS$_{2}$ TC-FET samples at $V_{ds}$ = +2 V. S1-S5 denotes the index of each sample.}
\label{fig:figS7}
\end{figure*}

We find that the value of tunneling current is dependent on the area of tunnel contacts, as shown in Fig. S6. The areas of left and right tunnel contacts are about 4 $\mu$m$^{2}$ and 1 $\mu$m$^{2}$, respectively (Fig. S6a). In the positive gate voltage, +6V, for example, the tunneling currents of 4 $\mu$m$^{2}$ tunnel contact is 4 nA (the blue curve in Fig. S6b), about 20 times larger compared to the 1 $\mu$m$^{2}$ tunnel contact which is 0.2 nA. On the other side of gate doping, -4V for example, the tunneling currents of 4 $\mu$m$^{2}$ tunnel contact is 40 times of the 1 $\mu$m$^{2}$ tunnel contacts, which is 12 nA and 0.3 nA, respectively.

\subsection{\normalsize 5. Cut-off frequency of the MoS$_{2}$ tunnel FETs.}

\begin{figure*}[h!]
\includegraphics[width=0.65\linewidth]{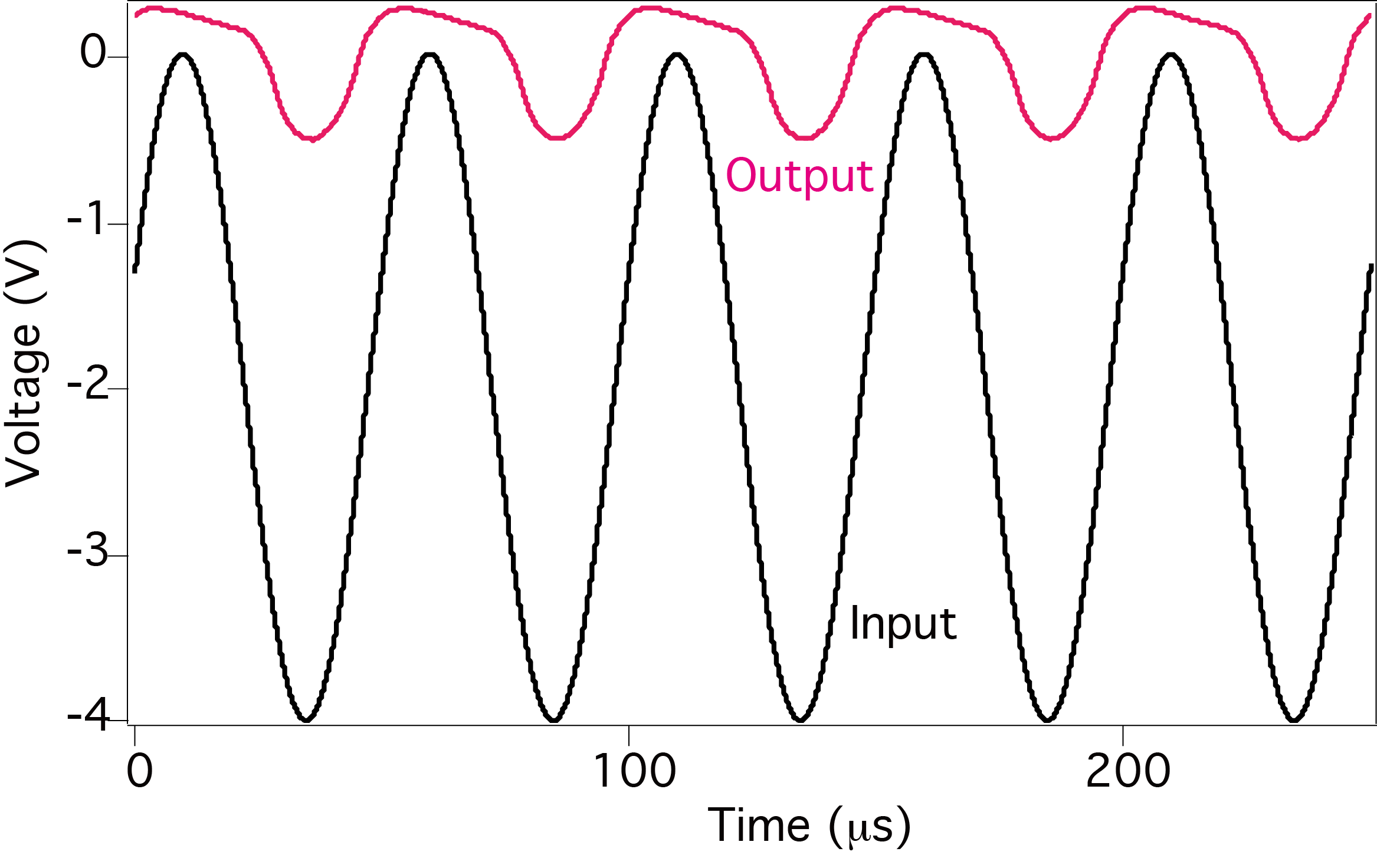}
\caption{When the input wave frequency is at the order of 20 kHz, output curve start to be distorted. In the mean time, de-phasing between input and put waves also start to show above 20 kHz. We then define 20 kHz as a cut-off frequency of our MoS$_{2}$ tunnel devices.}
\label{fig:figS8}
\end{figure*}

\subsection{\normalsize 6. First principles simulations of normal- and tunnel-contacted MoS$_{2}$ FETs.}

\begin{figure*}[h!]
\includegraphics[width=0.9\linewidth]{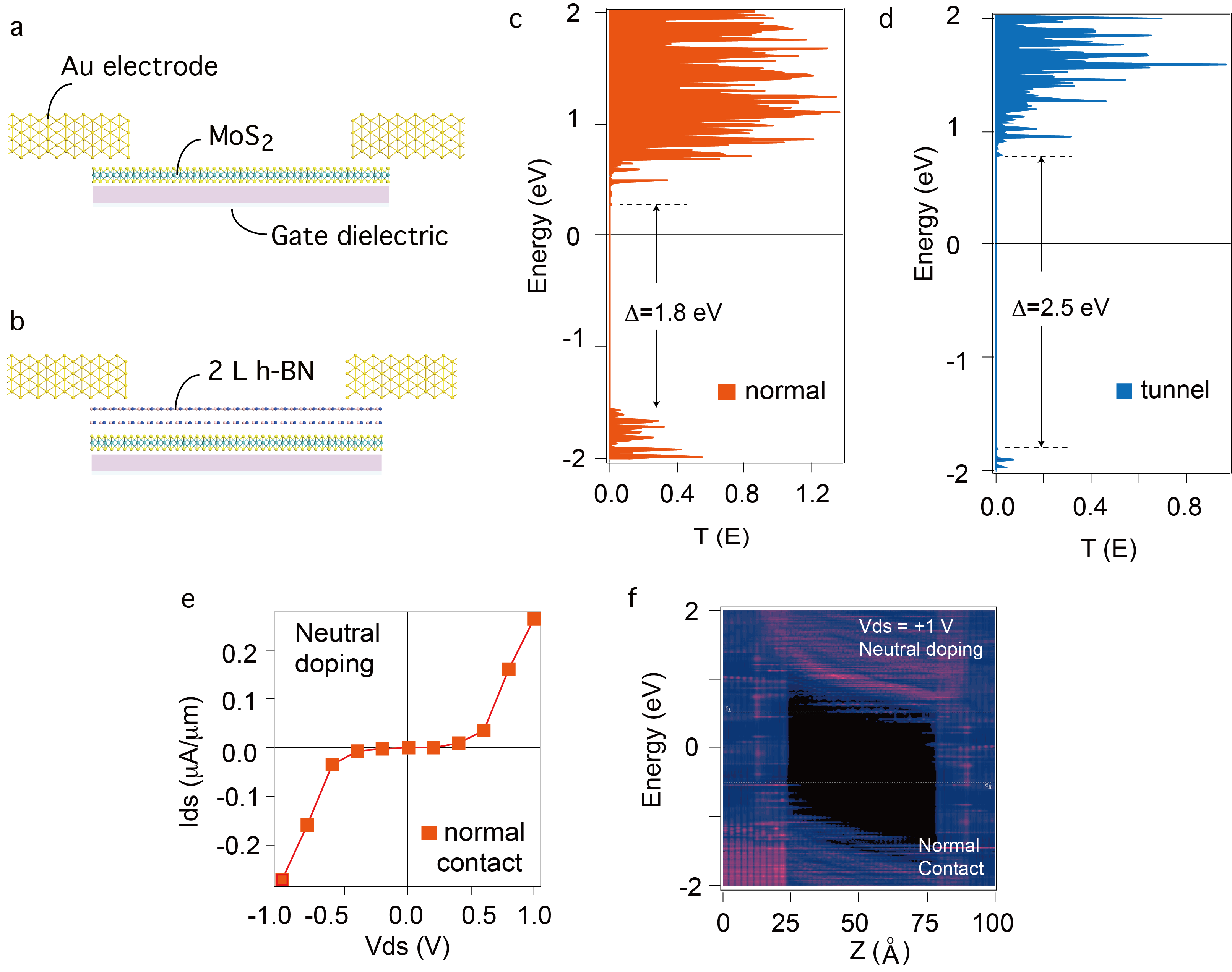}
\caption{(a)-(b) First principles simulation models of Au-contacted and 2L h-BN spaced devices. Channel length is set to be around 6 nm monolayer MoS$_{2}$. Back gate is replaced by an uniform dielectric with dielectric constant of 7.1, and thickness of 5 nm. (c) For normal contact device, one can see there is a transmission forbidden region (1.8 eV) from -1.56 eV to 0.24 eV which attributes to the band gap of monolayer MoS$_{2}$. Because the conduction band minimum (CBM) of MoS$_{2}$ is closer to Fermi level than that of valence band maximum (VBM), typical transport behaviour of $n$-type FET can be seen in this normal device (shown in (e)). (d) shows that when the monolayer MoS$_{2}$ is covered by 2-layered h-BN, the transmission forbidden region in Fig. 3b extends to 2.5 eV due to the associated effect of monolayer MoS$_{2}$ and 2-layered h-BN. The projected local density states (PLDOS) at $V_{ds}$=+1.0 V and at neutral doping in normal-contacted device is shown in Fig. S9f. It is clear seen that, different from those shown in the main text, the normal-contacted devices are subjected to the conventional picture of Fermi level pinning, and the DOS bending, where a strong Schottky barrier plays a significant role in the electron transport. Even though the model of simulation is simplified compared to real experimental conditions, the results listed here qualitatively support the observed phenomena of rectifying behavior in tunnel-contacted MoS$_{2}$ FETs.}
\label{fig:figS9}
\end{figure*}

\end{document}